\documentstyle[prl,aps,multicol,psfig,eqsecnum]{revtex}

\begin{document}

\preprint{adcode-quantphys.tex}

\title{Bosonic Quantum Codes for Amplitude Damping}

\author{Isaac L. Chuang$^{1,2}$, Debbie W. Leung$^{1}$, {\em and}\/
		Yoshihisa Yamamoto${^1}$}

\address{\vspace*{1.2ex}
 	\hspace*{0.5ex}{$^1$ERATO Quantum Fluctuation Project} \\
 	Edward L. Ginzton Laboratory, Stanford University, 
 		Stanford, CA 94305 \\[1.2ex] 
 	$^2$ {Institute for Theoretical Physics \\
	University of California, Santa Barbara, CA 93106 }}

\date{\today}
\maketitle


\begin{abstract}
Traditional quantum error correction involves the redundant encoding
of $k$ quantum bits using $n$ quantum bits to allow the detection and
correction of any $t$ bit error.  The smallest general $t=1$ code
requires $n=5$ for $k=1$.  However, the dominant error process in a
physical system is often well known, thus inviting the question: given
a specific error model, can more efficient codes be devised?  We
demonstrate new codes which correct just amplitude damping errors
which allow, for example, a $t=1$, $k=1$ code using effectively
$n=4.6$.  Our scheme is based on using bosonic states of photons in a
finite number of optical modes.  We present necessary and sufficient
conditions for the codes, and describe construction algorithms, 
physical implementation, and performance bounds. 
\end{abstract}


\begin{multicols}{2}[]

\section{Introduction}

Information is classically measured in units of bits, which are deterministic
two-state systems that are said to exist either as logical zero, $0_L$ or
logical one $1_L$.  However, such a representation is only an approximation of
reality, which is described at small size scales by the laws of quantum
physics.  Quantum-mechanical two-state systems (``qubits'') differ from the
classical bit in that they may exist in a superposition of the two states, for
example as $|{\psi}\rangle = a|{0_L}\rangle + b|{1_L}\rangle$, where $a$ and
$b$ are arbitrary complex coefficients which satisfy $|a|^2 + |b|^2=1$.  Two
continuous real parameters are needed to describe the state of a qubit, and in
this sense, more information is somehow carried in it than by a classical bit.
Furthermore, qubits may not be cloned\cite{Wooters82}, and even more
importantly, they may exist in {\em entangled states} where, for example, two
qubits only carry one qubit of {\em quantum information}.

Unfortunately, quantum information is (partially) lost whenever a quantum
system is observed (whether deliberately or inadvertently).  This {\em
decoherence} process plays a role analogous to noise in a classical
communication channel.  A major advance in quantum information theory has been
the discovery that quantum information can be redundantly encoded in such a
manner that it may be efficiently transmitted with arbitrarily high fidelity
through a decohering quantum channel.  The {\em quantum error correction
codes}\cite{Shor95b,Steane96} which make this possible are analogous to
classical codes for binary memoryless channels.  Corresponding codes for
classical linear codes and Reed-Muller codes have been
found\cite{Caldebank95,Steane96a}.

These quantum coding schemes are based on a model for the decoherence of
qubits, in which three kinds of errors can occur: bit flips
($|{0}\rangle\leftrightarrow|{1}\rangle$), phase flips
($|{1}\rangle\leftrightarrow -|{1}\rangle$), and both simultaneously.  This
model is general; it describes all possible decoherence mechanisms for a
qubit.  However, in a given physical system, the dominant decoherence process
is of a specific nature which may admit a simpler description.  For example,
in phase damping, no bit flips occur.  The question therefore arises: given a
particular decoherence process, what is the optimal quantum error correction
scheme?

We do not yet know how to handle this general problem.  However in this paper,
we report on progress towards a solution, by demonstrating a new class of
quantum error correction codes which correct only one particular decoherence
process known as amplitude damping.  Our approach is similar in philosophy to
that of \cite{Plenio96a}, but in contrast to that and other previous schemes,
instead of qubits, which live in a two-dimensional Hilbert space, we utilize
bosonic systems which occupy the Hilbert space $|{0}\rangle\cdots
|{N}\rangle$.  We are unaware of any classical analog to our codes.  We
present possible physical implementations of our scheme, and conclude with a
comparison with existing binary codes.

\section{Amplitude Damping Model}

Noise is a fundamental process which accompanies the dynamics of any open
system.  Traditionally, the dynamics of an open quantum system are described
by a ``master equation'' \cite{Louisell}.  The system $a$, described initially
by the density matrix $\rho_{a0}$, couples to a bath $b$ through an
interaction Hamiltonian $H_I$.  Evolution generates an entangled state of the
total system $\rho$, and we trace over the bath state at the end.  Taking the
system to be a simple harmonic oscillator, and using the simplest bilinear
interaction in which photons are exchanged back and forth with the bath,
we find in the Born and Markov approximations an equation of motion for the
density matrix
\begin{equation}
	\dot{\rho} = -\frac{\lambda}{2}({a^\dagger a}\rho + \rho{a^\dagger a} 
		- 2a\rho a^\dagger)
\,,
\label{eq:adeqmot}
\end{equation}
where $\lambda$ is a coupling constant between the system and the
environment. We have set $\langle {b^\dagger b}\rangle =0$ in
(\ref{eq:adeqmot}) to reflect a reservoir at temperature $kT$ much smaller
than the system's energy scale $\hbar\omega$.  This process describes the
gradual loss of energy from the system to the environment, and is known as
{\em amplitude damping}\cite{Gardiner91,Louisell}.

Mathematically, the evolution of the density matrix from $t$ to $t'$ in a
particular process may be described as a linear transformation from one
density matrix $\rho$ to another, $\rho'$.  This may be expressed (in the
``positive operator sum representation'') as
\begin{equation}
	\rho' = \sum_k A_k \rho A^\dagger_k
\,,
\end{equation}
where $A_k$ are positive linear operators, sometimes referred to as ``Krauss
effects\cite{Kraus83a},'' which are related to the Lindblad operators
appearing in Eq.(\ref{eq:adeqmot}).  For amplitude damping, we find that
\begin{equation}
	A_k = 
	_b\langle{k}| e^{\chi (a^\dagger b - {b^\dagger} a)} |{0}\rangle _b
\,,
\end{equation}
where $\gamma = 1-\cos^2\chi$ is the probability of loosing a single photon
from the system between the initial and final times.  After tracing over the
final environmental state, we find the operators in the Hilbert space of the
system alone:
\begin{equation}
	A_k = \sqrt{\left(\begin{array}{r}{n}\\{k}\end{array}\right)}\,
		\sqrt{(1-\gamma)^{n-k} \gamma^k} \, 
		|{n-k}\rangle\langle{n}|
\,.
\label{eq:adak}
\end{equation}

If the initial state is pure, it may be written as $\rho =
|{\psi}\rangle\langle{\psi}|$.  The final state $\rho'$ may be elegantly
described as an explicit mixture of pure states given by
\begin{equation}
	[ \psi' \rangle  = \bigoplus_{k=0}^{N} A_k |{\psi}\rangle
\,,
\end{equation}
where $N$ is the maximum occupation number of a single bosonic mode.  Here,
the ``$\oplus$'' symbol represents a tensor sum of states, and $[\psi'\rangle
$ is a convenient shorthand used to denote a mixed state, as distinguished
from a pure state $|\psi\rangle $.  In other words,
\begin{equation}
	\rho' = [\psi' \rangle \langle  \psi' ]
	= \sum_{k=0}^N A_k |\psi\rangle  \langle  \psi| A_k^\dagger
\,.
\end{equation}
The mixed state $[\psi'\rangle $ is a tensor sum of $N+1$ (un-normalized) pure
states which describe the $N+1$ possible final states of the system; one may
interpret these as non-interfering ``alternative
histories\cite{Gell-Mann93}.''  The normalization of each pure state gives its
probability of occurrence.  In general, $k$ describes the number of photons
lost to the environment.  It is important that even when no photons are lost
to the environment, then the state of the system is changed.  

So far, we have described the effect of amplitude damping on a single
mode system.  Consider now a system with $m$ modes, and let us use
$A_{kj}$ to denote the action of the effect $A_k$ on the $j^{th}$ mode
of a state, $j\in[1,m]$.  After amplitude damping, the initial pure state
\begin{equation}
	|{\psi_{in}}\rangle = | n_1\ldots n_m \rangle 
\end{equation}
becomes the mixed state
\begin{equation}
	[\psi_{out} \rangle  
	= \left[\rule{0pt}{2.4ex} \bigoplus_{k=0}^{N} A_{k0} |{n_0}\rangle
		\right] 
		\cdots
		\left[\rule{0pt}{2.4ex} {\bigoplus_{k=0}^{N} A_{km}
			|{n_m}\rangle 
			}\right] 
\,,
\label{eq:mmodead}
\end{equation}
where there are now $(N+1)^m$ possible final states.  It is convenient
to use the shorthand notation
\begin{equation}
	A_{\tilde{k}} = A_{k_00} \cdots A_{k_mm}
\,,
\end{equation}
where $k_j$ is the $j^{th}$ digit of the number $\tilde{k}$ written in base
N+1, so that we may rewrite Eq.(\ref{eq:mmodead}) as
\begin{equation}
	[\psi_{out} \rangle  = \bigoplus_{\tilde{k}=0}^{(N+1)^m-1} 
				A_{\tilde{k}} |{\psi_{in}}\rangle
\,.
\end{equation}
Note that identical states in a tensor sum can be combined using the
rule
\begin{equation}
	a [\psi\rangle  \oplus b [\psi\rangle  = \sqrt{|a|^2 + |b|^2}
		[\psi\rangle  
\,,
\end{equation}
since an overall phase doesn't matter (assuming no entanglement with
other systems).  

As an example, amplitude damping of the state
\begin{equation}
	|\psi_{in}\rangle  = a |{01}\rangle+ b |{10}\rangle
\label{eq:dualrail}
\end{equation}
gives, using
\begin{eqnarray}
	A_0 &=& |0\rangle \langle 0| + \sqrt{1-\gamma} |1\rangle \langle 1|
\label{eq:adqubit}
\\	A_1 &=& \sqrt{\gamma} |0\rangle \langle 1|
\,,
\label{eq:adqubitend}
\end{eqnarray}
the output state
\begin{eqnarray}
	[\psi_{out}\rangle  
	&=& A_{00} |\psi_{in}\rangle  \oplus A_{01} |\psi_{in}\rangle 
		\oplus A_{10} |\psi_{in}\rangle  \oplus A_{11}
			|\psi_{in}\rangle  
\\
	&=& \sqrt{1-\gamma} |\psi_{in}\rangle  \oplus \sqrt{\gamma}
		|{00}\rangle 
\,.
\end{eqnarray}
This result can be understood intuitively: the original state only
contains a single photon, and thus, whenever it is lost, the final
state must be the vacuum.  This example indicates that the state of
Eq.(\ref{eq:dualrail}) is useful for detection of a single photon
loss.  However, no useful information about $a$ and $b$ can be
extracted from the vacuum state, and so it is not useful for error
{\em correction}.

\section{Examples}

Let us motivate the remainder of this paper by considering the following
example: We {\em encode} the logical zero and one states of a single qubit as
\begin{equation}
	|0_{L}\rangle  = \left[\rule{0pt}{2.4ex}{
		\frac{|{40}\rangle+|{04}\rangle}{\sqrt{2}} 
			}\right] 
\hspace*{3ex}
	|1_{L}\rangle  = |{22}\rangle
\,,
\end{equation}
such that the initial state is the arbitrary qubit 
\begin{equation}
	|\psi_{in}\rangle  = a |0_L\rangle  + b |1_L\rangle 
\,.
\end{equation}
The possible outcomes after amplitude damping may be written as
\begin{equation}
	[\psi_{out}\rangle  = \bigoplus_{\tilde{k}} |\phi_{\tilde{k}}\rangle 
	= \bigoplus_{\tilde{k}} A_{\tilde{k}} |\psi_{in}\rangle 
\,,
\end{equation}
where we shall express $\tilde{k}$ as a base 5 numeral, and
$|\phi_{\tilde{k}}\rangle $ is an unnormalized pure state (the norm of which
gives its probability for occuring in the mixture).  For small loss
probability $\gamma$, the most likely final state will be
\begin{equation}
	|\phi_{00}\rangle  = (1-\gamma)^{2} |\psi_{in}\rangle 
\,,
\end{equation}
corresponding to no quanta being lost to the bath.
The next most likely states result from the loss of a single photon:
\begin{eqnarray}
	|\phi_{01}\rangle  &=& \sqrt{2\gamma} (1-\gamma)^{3/2}
				\left[\rule{0pt}{2.4ex}{ a |03\rangle  + b
				|{21}\rangle }\right] 
\\
	|\phi_{10}\rangle  &=& \sqrt{2\gamma} (1-\gamma)^{3/2}
				\left[\rule{0pt}{2.4ex}{ a |30\rangle  + b
				|{12}\rangle }\right] 
\,.
\end{eqnarray}
States resulting from the loss of more than one quantum occur with
probabilities of order $\gamma^{2}$.  Therefore, we limit our correction
scheme to errors losing at most one quantum. Each such error $E_{i}$ takes
$|0_{L}\rangle $ and $|1_{L}\rangle $ to states $|0_{L}\rangle _{i}$ and
$|1_{L}\rangle _{i}$ respectively.  The key is that $|0_{L}\rangle $,
$|1_{L}\rangle $, $|0_{L}\rangle _{i}$ and $|1_{L}\rangle _{i}$ $\forall i$
are mutually orthogonal, and so are $|\phi_{00}\rangle $, $|\phi_{01}\rangle
$, and $|\phi_{10}\rangle $. In principle, a (``quantum non-demolition'')
measurement scheme can detect all error syndromes.  Furthermore, for each $i$,
the norms of $|0_{L}\rangle _{i}$ and $|1_{L}\rangle _{i}$ are equal.  After
detecting an error syndrome, one can apply an appropriate unitary
transformation converting $|0_{L}\rangle _{i}$ and $|1_{L}\rangle _{i}$, to
$|0_{L}\rangle $ and $|1_{L}\rangle $ respectively.  This makes possible the
correction:
\begin{equation}
	a |0_{L}\rangle _{i} + b |1_{L}\rangle _{i}
\rightarrow
	\alpha \left[{a |0_{L}\rangle  +  b |1_{L}\rangle }\right]
\,,
\end{equation}
where $\alpha$ is independent of $a$, $b$.  Note, this is done without any
information about $a$, $b$, and without diminishing the amplitude of the
erroneous state.  For this particular code, the output state has fidelity (see
Eq.(\ref{eq:fido})\cite{Schum95}) ${\cal F} = 1-6\gamma^2$ with respect to
the input.

As a comparison, consider the code:
\begin{equation}
	|0_{L}\rangle =|11\rangle ,~~~ |1_{L}\rangle =|22\rangle ,
\end{equation}
with the most probable state:
\begin{equation}
	|\phi_{00}\rangle  = a (1-\gamma) |11\rangle  + b  (1-\gamma)^{2}
				|22\rangle  
\,.
\end{equation}
No unitary transformation will bring it back to 
\begin{equation}
	a |11\rangle  + b |22\rangle 
\,.
\end{equation}
unless $a$, $b$ is predetermined (a non-unitary transformation can revert the
change, but it will reduce the fidelity of the correction process to 1-${\cal
O}(\gamma)$).

In the remainder of the paper, we shall describe the criteria for a scheme 
in which $k$ qubits may be encoded so that loss up to $t$ quanta may be
corrected.  For small $t$, a scheme will be exhibited.

\section{Code Criteria}

Quantum error correction is just the reversing of some effect due to
decoherence.  General criteria for this to be possible have been given in the
literature\cite{Nielsen96c}.  In this particular case, we may express the
required conditions in the following manner.  Let $\{|c_0\rangle \cdots
|c_l\rangle \cdots |c_{l_o}\rangle \}$ be $l_o+1$ codewords which encode
orthogonal logical states within the $m$ mode Hilbert space with maximum total
photon number $N$, and define ${\cal K}(t)$ as the set of all $m$-digit base
$N+1$ numbers whose digits sum to $t$ (corresponding to $t$ errors).  The
logical states must satisfy
\begin{eqnarray}
	\langle  c_{l_1} | A^\dagger_{\tilde{k}} A_{\tilde{k}'} |c_{l_2}
				\rangle  &=& 0 
	\hfill\hspace*{7ex}\mbox{for $l_1 \neq l_2$ or 
	$\tilde{k}\neq \tilde{k}'$}
\label{eq:critone}
\\
	\langle  c_l | A^\dagger_{\tilde{k}} A_{\tilde{k}} |c_l \rangle 
	&=& g_{\tilde{k}}
	\hfill\hspace*{6ex}\mbox{$\forall i$}
\label{eq:crittwo}
\end{eqnarray}
for all $\tilde{k}, \tilde{k}' \in \bigcup_{s\leq t}{\cal K}(s)$.  
Here, $g_{\tilde{k}}$ is
some constant which depends only on $\tilde{k}$.  This is an extension of the
non-degenerate quantum error correction code criteria given by \cite{Knill96}
to the case where none of the $A_k$ operators are identity.  The first
condition requires that all erroneous states be orthogonal, and the second
requires that the encoded Hilbert space not be deformed.  Here, we present an
explicit statement of these two conditions as algebraic conditions on the code
construction.

We consider first a codeword $|c_l\rangle $ which may be expressed as an
equally weighted sum of $N_l$ energy eigenstates, $|n_1\ldots n_m\rangle $ (we
shall refer to these as {\em quasi-classical states}, ``QCS'' for short).
When all the QCS are equally weighted, we call the code ``balanced''.
Otherwise, the code is referred to as ``unbalanced''.  Each codeword can be
represented by a matrix with $m$ columns and $N_l$ rows, each row being one of
the QCS in the codeword.  For instance, if
\begin{equation}
	|c_l\rangle =\frac{1}{\sqrt{N_l}} \left[\rule{0pt}{2.4ex}{
		|n_{11}\ldots n_{1m}\rangle +\ldots+|n_{N_l1}\ldots
			n_{N_lm}\rangle  
	}\right]
\,,
\end{equation}
then the corresponding matrix ${\cal M}_l$ is:
\begin{equation}
	\left[
	\begin{array}{clcr}
		n_{11} & n_{12} & \ldots & n_{1m} \\
		n_{21} & n_{22} & \ldots & n_{2m} \\
		\ldots & \ldots & \ldots \ldots & \\
		n_{N_l1} & n_{N_l2} & \ldots & n_{N_lm}
	\end{array}
	\right] 
\,.
\end{equation}

For $t=0$ errors, we have ${\cal K}(0) = \{0\}$, and $A_0|n_{i1}\ldots
n_{im}\rangle = (1-\gamma)^{RS_i/2} |n_{i1}\ldots n_{im}\rangle $ where the
row sum $RS_i=\sum_{j=1}^{m} n_{ij}$.  Criteria Eq.(\ref{eq:crittwo}) requires
the amplitudes of $A_0|c_l\rangle $ be the same for all $|c_l\rangle $, that
is:
\begin{equation}
	\frac{1}{N_l}\sum_{i=1}^{N_l} (1-\gamma)^{RS_i/2} 
\,
\label{eq:rowsum}
\end{equation}
be the same for all $|c_l\rangle $. A sufficient condition for this is the
equality of all the $RS_i$ for all $i$ and for all codewords $|c_l\rangle $.
Alternatively, one may say that the sum of any row from any ${\cal M}_l$
equals $N$, the same total photon number in all the QCS.  Denote the set of
QCS with $m$ modes and total photon number $N$ as ${\cal Q}(N,m)$.  It follows
that if we construct all the codewords from states in ${\cal Q}(N,m)$, then
the non-deformation constraint Eq.(\ref{eq:crittwo}) is satisfied for $t=0$.
Physically, this requirement stems from the fact that a state with higher
number of quanta decays faster.  To preserve the a posteriori probabilities of
each codeword, we must encode them in a subspace in which the decay
probabilities are equal for all of them.

Similarly, for $t=1$ errors, we have ${\cal K}(1) = \{0\cdots 1, 0\cdots 10,
\ldots, 1\cdots0 \}$, and, for example,
\begin{eqnarray}
	A_{0\cdots 1} |c_l\rangle 
	&=& A_{0\cdots 1} 
		\frac{1}{\sqrt{N_l}} \sum_{i=1}^{N_l} |n_{i1}\ldots
			n_{im}\rangle  
\\	&=& \sum_{i=1}^{N_l} \sqrt{\frac{n_{im}\gamma(1-\gamma)^{N-1}}
				      {N_l}}
			   |n_{i1}\ldots n_{im}-1\rangle 
\,,
\end{eqnarray}
and
\begin{equation}
	\langle c_l| A_{0\cdots 1}^\dagger A_{0\cdots 1} |c_l\rangle 
	= \frac{\gamma(1-\gamma)^{N-1}}{N_l} \sum_{i=1}^{N_l} n_{im}
\,.
\end{equation}
The non-deformation criteria requires the above sum to be the same for all
codewords.  Equivalently, the column sum of the $m^{th}$ column of each
codeword divided by $N_l$ has to be independent of the codeword.  Similar
expressions for other $A_{\tilde{k}}$ give rise to similar criteria for each
column separately.

We therefore have the following:
\begin{quote}
	{\em Lemma~1.1}: Let each codeword be expressed as an $m$ column,
	$N_l$ row matrix with elements $n_{ij}$.  If we choose codewords such
	that $\sum_i n_{ij}/N_l = y_j$ for all $|c_l\rangle $, then $\langle
	c_l| A^\dagger_{\tilde{k}} A_{\tilde{k}} |c_l\rangle = g_{\tilde{k}}$,
	$\forall \tilde{k} \in {\cal K}(1)$.
\\[1.5ex]
	{\em Proof:} As above. $\Box$
\end{quote}
This criteria corresponds to certain symmetry requirement among the various
codewords.  For $t=2$, ${\cal K}(2)=\{0\cdots 02,0\cdots 11,0\cdots 20
\ldots, 20\cdots0 \}$. Working out $A_{\tilde{k}}|c_l\rangle $ for each
$\tilde{k}$ and applying the criteria for $t=0,1,2$, one arrives at the
following:
\begin{quote}
	{\em Lemma~1.2}: Same setting as $Lemma1.1$. Let us choose codewords
	which satisfy the non-deformation criteria for $t=1$, and such that
	$\sum_i n_{ij_1} n_{ij_2}/N_l = y_{j_1,j_2}$ for all $|c_l\rangle $,
	where $j_1,j_2\in[1,N_l]$, $j_1,j_2$ may or may not be distinct.  Then
	$\langle c_l| A^\dagger_{\tilde{k}} A_{\tilde{k}} |c_l\rangle =
	g_{\tilde{k}}$, $\forall \tilde{k} \in {\cal K}(2)$.
\\[1.5ex]
	{\em Proof:} See Appendix~A. $\Box$
\end{quote}
Generalization to arbitrary  $t$ is as follows:
\begin{quote}
	{\em Theorem~1}: Let each codeword be expressed as an $m$ column,
	$N_l$ row matrix with elements $n_{ij}$.  If we choose codewords such
	that $\sum_i n_{ij_1} n_{ij_2} \cdots n_{ij_t}/N_l =
	y_{j_1,j_2,\ldots,j_s}$ independent of $|c_l\rangle $, $\forall l$,
	$\forall (j_1,j_2,\ldots,j_s) \in [1,N_l]^s$ and $\forall s \in
	[1,t]$, then $\langle c_l| A^\dagger_{\tilde{k}} A_{\tilde{k}}
	|c_l\rangle = g_{\tilde{k}}$, $\forall \tilde{k} \in \bigcup_{s=1}^t
	{\cal K}(s)$, $\forall l$.
\\[1.5ex]
	{\em Proof:} See Appendix~A. $\Box$
\end{quote}
The above theorem can be generalized to unbalanced codes in which codewords
are unequally weighted superpositions of QCS.  If the amplitudes of the QCS in
$|c_l\rangle $ are $(\sqrt{\mu_1},\sqrt{\mu_2},\cdots,\sqrt{\mu_{N_l}})$, we
replace the sum $\sum_i n_{ij_1} n_{ij_2} \cdots n_{ij_t}/N_l$ by $\sum_i
\mu_i n_{ij_1} n_{ij_2} \cdots n_{ij_t}$, i.e., we replace the equal weights
$\frac{1}{N_l}$ by the $\mu_i$'s (the derivation of {\em Theorem 1} in the
unbalanced case is a straightforward generalization of the balanced case and
we will skip the proof).

As $t$ increases, the non-deformation criteria becomes very restrictive. We
have found unbalanced codes by numerical search correcting up to $t\leq4$, as
(Section~\ref{sec:moreexamples}) which have no analogues in the balanced
codes.  On the other hand, for $t \leq 2$, we found simple construction
algorithms for balanced codes with no apparent counterparts for the unbalanced
codes.

It should also be noted that the $t=0$ non-deformation criteria, that row sums
(total number of excitations in each QCS) be equal for all rows and for all
matrices (codewords), is not a necessary condition.  An example is Shor's
9-bit code\cite{Shor95b}:
\begin{eqnarray}
	&|0_L\rangle &=(|000\rangle +|111\rangle )^{\otimes3}
\\
	&|1_L\rangle &=(|000\rangle -|111\rangle )^{\otimes3}
\,.
\end{eqnarray}
The QCS have different number of $1$'s, but Eq.(\ref{eq:rowsum}) is equal for
$|0_L\rangle $ and $|1_L\rangle $. However, code criteria for $t>1$ will be
extremely complicated when row sums are different, and treatment of such codes
are outside the scope of the present discussion.

The other criteria, the orthogonality constraint Eq.(\ref{eq:critone}) can be
satisfied as follows.  Let $|u\rangle = |u_1\ldots u_m\rangle $ and $|v\rangle
= |v_1\ldots v_m\rangle $ be two states in ${\cal Q}(N,m)$.  We define the
{\em distance} between $u$ and $v$ as
\begin{equation}
	{\cal D}(u,v) = \frac{1}{2} \sum_i |u_i - v_i|
\,.
\end{equation}
Clearly, $0\leq {\cal D}\leq N$. Moreover, ${\cal D}(u,v) = {\cal D}(v,u)$, 
${\cal D}(u,u)=0$, and 
\begin{eqnarray}
	{\cal D}(u,v)+{\cal D}(v,w)&=&\frac{1}{2} \sum_i |u_i-v_i|+|v_i-w_i|
\\
                                &\geq&\frac{1}{2} \sum_i |u_i-w_i|
\\
                                   &=&{\cal D}(u,w)
\,.
\end{eqnarray}
Thus ${\cal D}$ is a metric on the discrete space ${\cal Q}(N,m)$.  (For
binary states, this is half of the Hamming distance).  Define the distance
between two codewords $|c_1\rangle $ and $|c_2\rangle $ to be the minimum of
${\cal D}(u_1,u_2)$ with $|u_1\rangle $, $|u_2\rangle $ being QCS in
$|c_1\rangle $ and $|c_2\rangle $ respectively. For codewords with
non-negative amplitudes of the constituent QCS, two codewords are orthogonal
{\em iff} their distance is non zero.  We therefore have the following:
\begin{quote}
	{\em Theorem~2}: Let $|c_1\rangle $ and $|c_2\rangle $ be two
	codewords formed from states in ${\cal Q}_1$ and ${\cal Q}_2$
	respectively, where ${\cal Q}_1, {\cal Q}_2 \subset {\cal Q}(N,m)$ and
	${\cal D}(u_1,u_2)> t$ $\forall u_1\in {\cal Q}_1, u_2\in{\cal Q}_2$.
	Then $\langle c_l| A^\dagger_{\tilde{k}} A_{\tilde{k}'} |c_2\rangle =
	0$, $\forall \tilde{k}, \tilde{k}' \in \bigcup_{s\leq t}{\cal K}(s)$.
\\[1.5ex]
	{\em Proof:} Let $A_{\tilde{k}'}|c_1\rangle =|d_1\rangle $,
	$A_{\tilde{k}}|c_2\rangle =|d_2\rangle $, and let $|v_1\rangle $,
	$|v_2\rangle $ be QCS in $|d_1\rangle $, $|d_2\rangle $ respectively
	$s.t.$ ${\cal D}(d_1,d_2) = {\cal D}(v_1,v_2)$.  Let $|u_1\rangle $,
	$|u_2\rangle $ be the original QCS in $|c_1\rangle $, $|c_2\rangle $
	before the error.  Then, ${\cal D}(u_1,v_1) = {\cal D}(u_2,v_2) \leq
	t/2$, and ${\cal D}(v_1,v_2) + {\cal D}(u_1,v_1) + {\cal D}(u_2,v_2)
	\geq {\cal D}(u_1,u_2) > t$. Hence, ${\cal D}(v_1,v_2) > 0$ and ${\cal
	D}(d_1,d_2) > 0$.  Therefore, $|d_1\rangle $ and $|d_2\rangle $ are
	orthogonal states. $\Box$
\end{quote}
In other words, by forming codewords using QCS which are sufficiently far
apart, then the orthogonality conditions is easily satisfied.

\section{Existence of codes}  

\label{sec:existence}

How large must $N$, $m$, and $N_l$ be to satisfy both the non-deformation
constraint, Eq.(\ref{eq:crittwo}), and the orthogonality constraint,
Eq.(\ref{eq:critone})?  We now show that an unbalanced code exists for
arbitrarily large $t$ if $N$ is allowed to be arbitrarily large, and give an
upper bound for the required $N$.  

Let $|c_0\rangle $, $|c_1\rangle $, $\ldots$, $|c_l\rangle $, $\ldots$,
$|c_{l_o}\rangle $, be $l_o+1$ codewords, each being an unequally-weighted
superposition of $N_l$ QCS in ${\cal Q}(n,m)$.  For convenience, define
\begin{equation}
	{\cal P}(n,m) = {\cal C}(n+m-1,m-1) \equiv \frac{(n+m-1)!}{n!(m-1)!}
\end{equation}
as the number of all possible partitions of the integer $n$, i.e., the number
of $x_i$ such that $x_1+x_2+\ldots+x_m=n$ \cite{Hardy60}.  Then if we choose
$N=nd$ such that
\begin{equation}
	{\cal P}(n,m) \geq N_1 + N_2 + \cdots + N_{l_o} = N_{QCS}
\,,
\end{equation}
where $N_{QCS}$ is the total number of QCS in the codewords, then by {\em
Theorem~2}, all the QCS involved can be chosen to be distinct, and
multiplication of the number states by $d=t+1$ allows the orthogonality
condition to be satisfied.

On the other hand, the non-deformation condition involves satisfying a certain
number of constraint equations, given by the total number of possible errors
times $l_o$. The number of errors involving losing $s$ photons from $m$ modes
is just the number of partitions of $s$ into $m$ parts, ${\cal P}(s,m)$.  Take
the QCS to be arbitrary, and solve the non-deformation constraint equations
(of {\em Theorem~1}, generalized to include unbalanced codes) as linear
equations for the weights of the QCS.  As long as the number of variables
($N_{QCS}$) are no fewer than the number of equations, solutions always exist.
We may also augment the system of equations by $l_o+1$ equations to ensure the
correct normalization of each codeword.  Hence, for $N$ which satisfy
\begin{eqnarray}
	1 + l_o + l_o \sum_{s=0}^t {\cal P}(s,m) 
	\leq N_{QCS}  
	\leq  {\cal P}(N/(t+1),m)
\label{eq:nqcs} 
\,,
\end{eqnarray}
codes with $m$ modes correcting $t$ errors exist.

For example, when $m=2$, Eq.(\ref{eq:nqcs}) becomes 
\begin{equation}
	\frac{l_o(t+2)(t+1)}{2}+l_0 + 1 \leq \frac{N}{t+1}+1 
\,,
\end{equation}
which gives a scaling law $N \approx t^3 l_o /2$.  Note that this upper bound
is generally much larger than necessary, as can be seen in the examples for
$t=3$ or $t=4$.  Much more efficient codes may be obtained, because the QCS
may be chosen to give redundant constraint equations.  This may be
accomplished either systematically (next section), or by numerical search
(Section~\ref{sec:moreexamples}).

\section{Construction Algorithm For $\mbox{$t$} \leq 2$ Balanced Codes}

So far, we have established criteria for and proved the existence of bosonic
codes for amplitude damping.  We now present an explicit procedure which
obtain a class of balanced codes to correct for $t=1$ and $t=2$ errors.

To correct $t=1$ error, consider ordered $m$-tuples $(x_1,x_2,\ldots,x_m)$
such that $x_1+x_2+\ldots+x_m=n$. We will use the same symbol ${\cal Q}(n,m)$
for the space of all such $m$-tuples as well as the space of all QCS $\{|x_1
x_2 \ldots x_m\rangle \}$.  We define an operation, ${\cal R}$ on ${\cal
Q}(n,m)$ which takes $(x_1,x_2,\ldots,x_m)$ to $(x_2,\ldots,x_m,x_1)$, i.e.,
the symbols are cyclically permuted. Define the order of an element to be the
size of its orbit under ${\cal R}$. It follows that the order $p$ must divide
m, and let $m=pq$.  An element of order $p$ looks like
$(x_1,\ldots,x_p,x_1,\ldots,x_p,\ldots,x_1,\ldots,x_p)$ with the string
$(x_1,\ldots,x_p)$ repeated $q$ times. The orbit looks like:
\begin{eqnarray}
	&&(x_1,x_2,\ldots,x_p,\ldots \ldots,x_1,x_2,\ldots,x_p)\\
	&&(x_2,x_3,\ldots,x_1,\ldots \ldots,x_2,x_3,\ldots,x_1)\\
	&&\cdots \\
	&&(x_p,x_1,\ldots,x_{p-1},\ldots \ldots,x_{p},x_1,\ldots,x_{p-1})
\,.
\end{eqnarray}

We form states by taking equal-weight superposition of QCS in each orbit:
\begin{eqnarray}
	|c\rangle  =& \frac{1}{\sqrt{p}}(&|x_1 x_2 \cdots x_p \cdots \cdots 
	x_1 x_2 \cdots x_p\rangle  \\
	 & + &|x_2 x_3 \cdots x_1 \cdots \cdots x_2 x_3 \cdots x_1\rangle  \\
	 & + &\cdots \\
	 & + &|x_p x_1 \cdots x_{p-1} \cdots \cdots x_p x_1 \cdots x_{p-1}
								\rangle )
\,.
\end{eqnarray}

States formed by distinct orbits are orthogonal, as the orbits partition
${\cal Q}(n,m)$.  Furthermore, we multiply each number in the QCS by $d$.  The
minimal separation of distinct QCS will be at least $d$, since distances come
as multiples of $d$ only. Hence, all the codewords will remain orthogonal
after errors of $t<d$ occur.  codewords are now in the form:
\begin{eqnarray}
	|c\rangle  =& \frac{1}{\sqrt{p}}(&|dx_1 dx_2 \cdots dx_p \cdots dx_1
				dx_2 
 \cdots dx_p\rangle  \\
	 & + &|dx_2 dx_3 \cdots dx_1 \cdots dx_2 dx_3 \cdots dx_1\rangle  \\
	 & + &\cdots \\
	 & + &|dx_p dx_1 \cdots dx_{p-1} \cdots dx_p dx_1 \cdots dx_{p-1}
								\rangle )
\,.
\end{eqnarray}

For the non-deformation criteria, the row sum is $nd=N$ by construction.  The
column sum divided by the normalization factor squared is (by {\em Lemma 1.1})
\begin{equation}
	\frac{dx_1+\cdots+dx_p}{p} = \frac{d n}{m}
\,
\end{equation}
in any codeword, independent of the order of the constituent QCS. 
Codes in examples (1)-(3) in Section~\ref{sec:moreexamples} 
are constructed in this way.

To correct for $t=2$ errors, the $t=1$ criteria have to be satisfied as well.
We take a subset of the $t=1$ codewords which will survive the extra
non-deformation criteria for $t=2$.  We also replace $d=2$ by $d \geq 3$. For
$m > 2$, pairs of codewords in the form
\begin{eqnarray}
	|0_L\rangle  =& \frac{1}{\sqrt{m}}(&|dx_1 dx_2 \cdots dx_m\rangle  \\
	 & + &|dx_2 dx_3 \cdots dx_1\rangle  \\
	 & + &\cdots \\
	 & + &|dx_m dx_1\cdots dx_{m-1}\rangle )
\,
\end{eqnarray}

\begin{eqnarray}
	|1_L\rangle  =& \frac{1}{\sqrt{m}}(&|dx_m \cdots dx_2 dx_1\rangle  \\
	 & + &|dx_{m-1} \cdots dx_1 dx_m\rangle  \\
	 & + &\cdots \\
	 & + &|dx_1 dx_m \cdots dx_2\rangle ) 
\,,
\end{eqnarray}
will always satisfy the non-deformation criteria for $t=2$ (proof omitted).
Examples~(4) and (5) in Section~\ref{sec:moreexamples} are constructed in this
way. This encodes only one qubit; we are still looking for $t=2$ codes which
can encode more qubits.

For $t\geq3$, we performed a numerical search for special QCS in which the
system of linear equations for the weights is linearly dependent.  In the best
case, the number of linear equations to be solved can be much reduced.
Therefore we can find codewords involving fewer QCS, fewer number of modes and
fewer number of quanta.  Although encoding is certainly possible with a much
smaller Hilbert space, we have not found a systematic way to generate such QCS.
Codes correcting $t\leq4$ errors are exhibited in
Section~\ref{sec:moreexamples}.

\section{Rates and Fidelities}

The performance of these bosonic quantum codes can be characterized by their
{\em rate} -- number of qubits communicated per qubit transmitted, and by
their {\em fidelity} -- the worst-case qubit degradation after decoding and
correction.  We discuss these two measures here.

The rate $r$ is given by the ratio of the number of encoded qubits to the
maximum number of qubits that can be accommodated in our Hilbert space:
\begin{equation}
	r = \frac{k}{m\log_2 (nd+1)}
\,,
\end{equation}
where $2^k$=number of codewords, and $(nd+1)^m$ is the size of the Hilbert
space in our code.  The exact number of possible codewords depends on the
choice of $N$ and $m$ (we have worked out a counting scheme, but omit the
details here). However, the majority of the QCS have order $m$. Hence, to a
good approximation, the number of codewords obtained is:
\begin{equation}
	2^k=\frac{{\cal P}(n,m)}{m}
\,,
\label{eq:rate}
\end{equation}
For small $n$, codewords involving fewer than $m$ QCS allow slightly more
qubits to be encoded compared with (\ref{eq:rate}).  This small gain can be
important in applications such as key distribution in quantum cryptography.

We now turn to the code fidelity ${\cal F}$, which we desire to know as a
function of the parameters $N$, $m$ and $t$.  It is defined as
\begin{equation}
	{\cal F} = {\rm min}_{\psi_{in}}
		\sum_{\tilde{k} \in \bigcup_{s\in[1,t]}{\cal K}(s)}
		 \langle \psi_{in}| A^\dagger_{\tilde{k}} A_{\tilde{k}}
			|\psi_{in}\rangle  
\,.
\label{eq:fido}
\end{equation}
Let the input state be expressed as a sum of codewords, $|\psi_{in}\rangle =
\sum_l \alpha_l |c_l\rangle $.  Then using the orthogonality and
non-deformation conditions, we find that
\begin{equation}
	\langle \psi_{in}| A^\dagger_{\tilde{k}} A_{\tilde{k}}
			|\psi_{in}\rangle  
	= | \langle c_l |A_{\tilde{k}}^\dagger A_{\tilde{k}} | c_l\rangle  |
\,,
\end{equation}
with $| c_l\rangle$ any one of the codewords in the right hand side.
Now, if we write each codeword as
\begin{eqnarray}
	|c_l\rangle &=&\sqrt{\mu_1} |n_{11} n_{12} \cdots n_{1m}\rangle 
\\	&+&\sqrt{\mu_2} |n_{21} n_{22} \cdots n_{2m}\rangle 
\\	&+&\cdots
\\      &+&\sqrt{\mu_{N_l}} |n_{N_l1} n_{N_l2} \cdots n_{N_lm}\rangle 
\,,
\end{eqnarray}
then, for $\tilde{k}$ = $(k_1,k_2,\ldots,k_m)$ $\in$ ${\cal K}(s)$:
\begin{equation}
	| \langle c_l| A_{\tilde{k}}^\dagger A_{\tilde{k}} | c_l\rangle  |
	=(1-\gamma)^{N-s} \gamma^s \sum_{i=1}^{N_l}\mu_i 
	   {\cal C}(n_{ij},k_j) 
\,,
\label{eq:bincoeff}
\end{equation}
and using the following relation for binomial coefficients:
\begin{equation}
	{\cal C}(N,s)=\sum_{\tilde{k}\in{\cal K}(s)}
	\sum_{i=1}^{N_l}\mu_i {\cal C}(n_{i1},k_1){\cal C}(n_{i2},k_2)
	\cdots {\cal C}(n_{im},k_m)
\,,
\label{eq:bla}
\end{equation}
we find that the fidelity is
\begin{eqnarray}
	{\cal F} &=& \sum_{s=1}^{t}(1-\gamma)^{N-s}\gamma^{s}{\cal C}(N,s)
\\		 &=& 1-{\cal C}(N,t+1)\gamma^{t+1}+{\cal O}(\gamma^{t+2})
\label{eq:explicitfid}
\,.
\end{eqnarray}
This expression holds for balanced codes as well as unbalanced codes.  The
amazing feature is that given a code which satisfies the orthogonality and
non-deformation constraints, ${\cal F}$ is independent of $m$; it is
determined only by $N$ and $t$.

One should note that although codes can be constructed to correct an arbitrary
number of photon loss, the more errors one wishes to correct, the more photons
are required.  On the other hand, having larger photon number states means a
higher probability for the system as a whole to suffer loss of quanta.  These
two effects compete against each other to give an upper bound on the fidelity,
which we can estimate as follows.  Let $N$ the
required photon number, and $t$ be the total number of errors to be corrected.
As previously discussed, due to the constraint equations which hold for error
correction to be possible, the two parameters can be reduced to one degree
of freedom, in terms of which we may estimate the optimal achievable fidelity.
In terms of $t$, the optimum fidelity for fixed $\gamma$ is obtained by setting
\begin{equation}
	\frac{d}{dt} \ln(1-{\cal F}) = 0
\,.
\end{equation}
From Eq.(\ref{eq:explicitfid}), this gives to first order in $\gamma$
\begin{equation}
	\frac{1}{{\cal C}} 
	\frac{\partial{\cal C}}{\partial N}\frac{dN}{dt}
	+
	\frac{1}{{\cal C}} \frac{d{\cal C}}{dt} + \ln\gamma = 0
\,.
\end{equation}
where ${\cal C}$ is a short hand notation for ${\cal C}(N,t+1)$. Using the
Stirling approximation for the factorials in ${\cal C}$, we obtain
\begin{equation}
	\ln \left( \frac{N}{N-t-1} \right) \frac{dN}{dt} 
	+ \ln \left( \frac{N-t-1}{t+1} \right) + \ln\gamma = 0
\,.
\label{eq:dfdt}
\end{equation}
In general, $N$ is much larger than $t$, which allows further simplification
of Eq.(\ref{eq:dfdt}):
\begin{equation}
	\frac{dN}{dt} \frac{t}{N} - \ln \left( \frac{t}{N} \right) + \ln\gamma
	= 0
\,.
\end{equation}
Generally, $N$ will asymptotically follow a power scaling law in $t$, so we
may approximate $ N \approx f l_o t^\alpha $ where the prefactor $f$ and
exponent $\alpha$ are approximately constant.  Solving for the optimum $t$ we
find that
\begin{equation}
	t_{opt}	\approx \left( e^{-\alpha} /\gamma f l_0 \right)^{1/(\alpha-1)}
\,.
\end{equation}
Plugging back into the Eq.(\ref{eq:explicitfid}) gives us an estimate for the
optimal achievable fidelity.  We thus have a loose bound on the fidelity of a
bosonic code with arbitrary QCS.  There is no theoretical bound on the number
of correctable errors.

\section{Explicit Codes}

\label{sec:moreexamples}

Some explicit codes resulting from our work are presented here.  States are
given unnormalized when the normalization factor is common for all codewords.
Codes are specified as $[[N,m,2^k,d]]$, where $N$ is the total number of
excitations in the QCS, $m$ is the number of modes for each QCS, $2^k$ is the
number of codewords and $d$ is the minimal distance between codewords. The
fidelity of all the codes are given by ${\cal F} \approx 1-{\cal C}(N,t+1)
\gamma^{t+1}$.

Example (1) -- $[[4,2,2,2]]$, $n=2$, $t=1$, fidelity ${\cal F} \approx
1-6\gamma^2$:
\begin{eqnarray}
	|{0_L}\rangle &=& \frac{1}{\sqrt{2}} [|{40}\rangle+|{04}\rangle]
\\	|{1_L}\rangle &=& |{22}\rangle
\end{eqnarray}

Example (2) -- $[[12,3,10,2]]$, $n=6$, $t=1$, fidelity ${\cal F} \approx
1-66\gamma^2$, labels given in hexadecimal ($c=12$, $a=10$):
\begin{eqnarray}
	|c_1\rangle  &=& \frac{1}{\sqrt{3}}[|{00c}\rangle + |{c00}\rangle +
		|{0c0}\rangle] 
\\	|c_2\rangle  &=& \frac{1}{\sqrt{3}}[|{02a}\rangle + |{a02}\rangle +
		|{2a0}\rangle] 
\\	|c_3\rangle  &=& \frac{1}{\sqrt{3}}[|{048}\rangle + |{804}\rangle +
		|{480}\rangle] 
\\	|c_4\rangle  &=& \frac{1}{\sqrt{3}}[|{066}\rangle + |{606}\rangle +
		|{660}\rangle] 
\\	|c_5\rangle  &=& \frac{1}{\sqrt{3}}[|{084}\rangle + |{408}\rangle +
		|{840}\rangle] 
\\	|c_6\rangle  &=& \frac{1}{\sqrt{3}}[|{0a2}\rangle + |{20a}\rangle +
		|{a20}\rangle] 
\\	|c_7\rangle  &=& \frac{1}{\sqrt{3}}[|{228}\rangle + |{822}\rangle +
		|{282}\rangle] 
\\	|c_8\rangle  &=& \frac{1}{\sqrt{3}}[|{246}\rangle + |{624}\rangle +
		|{462}\rangle] 
\\	|c_9\rangle  &=& \frac{1}{\sqrt{3}}[|{264}\rangle + |{642}\rangle +
		|{264}\rangle] 
\\	|c_a\rangle  &=& |{444}\rangle
\end{eqnarray}

Example (3) -- $[[6,3,4,2]]$, $n=3=0+1+2$, $t=1$, fidelity ${\cal F} \approx
1-15 \gamma^2$:
\begin{eqnarray}
	|{c_1}\rangle &=& |{600}\rangle + |{060}\rangle + |{006}\rangle
\\	|{c_2}\rangle &=& |{420}\rangle + |{204}\rangle + |{042}\rangle
\\	|{c_3}\rangle &=& |{240}\rangle + |{402}\rangle + |{024}\rangle
\\	|{c_4}\rangle &=& |{222}\rangle
\,.
\end{eqnarray}

Example (4) -- $[[9,3,2,3]]$, $n=3$, $t=2$, fidelity ${\cal F} \approx 1-
84\gamma^3$: Note this code differs from the previous one from having $d=3$
instead of $d=2$.  We take only $|c_2\rangle $ and $|c_3\rangle $ as codewords.
\begin{eqnarray}
	|0_L\rangle  &=& |{306}\rangle + |{063}\rangle + |{630}\rangle
\\	|1_L\rangle  &=& |{036}\rangle + |{360}\rangle + |{603}\rangle
\end{eqnarray}

Exapmle (5) -- $[[6,4,2,2]]$, $n$=6=0+1+2+3, fidelity ${\cal F} \approx 1-15
\gamma^2$: The minimal distance between QCS is $d=2$.  However, the QCS are
not generated by multiplying each number by $d=2$.
\begin{eqnarray}
	|{0_L}\rangle &=& |{0321}\rangle + |{1032}\rangle + |{2103}\rangle +
		|{3210}\rangle 
\\	|{1_L}\rangle &=& |{0123}\rangle + |{1230}\rangle + |{2301}\rangle +
		|{3012}\rangle 
\,.
\end{eqnarray}

Example (6) -- $[[7,2,2,2]]$, fidelity ${\cal F} \approx 1-21 \gamma^2$:
The codewords are not formed by cyclic permutations of the QCS. Note that column
one and two have different column sums.
\begin{eqnarray}
	|{0_L}\rangle &=& |{70}\rangle+|{16}\rangle
\\	|{1_L}\rangle &=& |{52}\rangle+|{34}\rangle
\end{eqnarray}

Example (7) -- $[[9,2,2,3]]$, fidelity ${\cal F} \approx 1-84\gamma^3$:
Unbalanced code that will tolerate $t=2$ errors.  Note that one codeword is
formed from the other by reversing the order of the modes.  (this symmetry
between the two modes is a sufficient condition for balanced codes with
$t=2$, $m\geq3$).
\begin{eqnarray}
	|{0_L}\rangle &=& \frac{1}{2}|{90}\rangle+ \frac{\sqrt{3}}{2}
		|{36}\rangle 
\\	|{1_L}\rangle &=& \frac{1}{2}|{09}\rangle+ \frac{\sqrt{3}}{2}
		|{63}\rangle 
\end{eqnarray}

Example (8) -- $[[9,3,2,3]]$, fidelity ${\cal F} \approx 1-84\gamma^3$:
Unbalanced code that will tolerate $t=2$ errors, showing that the symmetry is
not a necessary condition for correcting $t=2$ errors.
\begin{eqnarray}
	|0_L\rangle &=&\frac{1}{\sqrt{3}}[|{036}\rangle + |{306}\rangle +
		|{360}\rangle] 
\\
	|1_L\rangle&=&\frac{1}{3}[\sqrt{6}|{333}\rangle+
			\sqrt{2}|{009}\rangle+|{090}\rangle]  
\end{eqnarray}

Example (9) -- $[[16,2,2,4]]$, fidelity ${\cal F} \approx 1-1820\gamma^4$:
Unbalanced code that will tolerate $t=3$ errors.  Labels are given in base
17. $c$ and $g$ denote 12 and 16 respectively.
\begin{eqnarray}
	|0_L\rangle &=&\frac{1}{\sqrt{8}}[|{0g}\rangle+
		|{g0}\rangle+\sqrt{6}|{88}\rangle]
\\
	|1_L\rangle &=&\frac{1}{\sqrt{2}}[|{4c}\rangle+|{c4}\rangle]
\end{eqnarray}

Example (10) -- $[[20,3,2,4]]$, fidelity ${\cal F} \approx 1-4845\gamma^4$:
Another unbalanced code that will tolerate $t=3$ errors. Labels are given in
base 21. $c$, $g$ and $k$ denote 12, 16, and 20 respectively.
\begin{eqnarray}
	|0_L\rangle &=&\frac{1}{5}[|{04g}\rangle+
		2 |{40g}\rangle+2 \sqrt{5}|{0k0}\rangle]
\\
	|1_L\rangle &=&\frac{1}{\sqrt{5}}[\sqrt{2}|{44c}\rangle+
		\sqrt{3}|{488}\rangle]
\end{eqnarray}

Example (11) -- $[[50,2,2,5]]$, fidelity ${\cal F} \approx 1-2118760\gamma^5$:
Note the rapid growth in the numerical factor in the second term.  To correct
for large number of errors, we need to encode a qubit in a large Hilbert
space, but emission probabilities are large for high number states.  This puts
a limit of performance in our codes.  The actual code involves numbers five
times the numbers shown below. $a$ denotes 10.
\begin{eqnarray}
	|0_L\rangle &=&\sqrt{\frac{1}{18}}|{0a}\rangle+
		\sqrt{\frac{5}{9}}|{46}\rangle
\\            &+&\sqrt{\frac{1}{3}}|{82}\rangle+\sqrt{\frac{2}{45}}|{91}\rangle
\\	|1_L\rangle &=&\sqrt{\frac{1}{18}}|{19}\rangle+
			\sqrt{\frac{1}{6}}|{28}\rangle
\\	      &+&\sqrt{\frac{33}{90}}|{55}\rangle+
			\sqrt{\frac{1}{3}}|{73}\rangle
			+\sqrt{\frac{7}{90}}|{a0}\rangle
\,.
\end{eqnarray}
%

\section{Physical Implementation}

Encoding and decoding of the codes we have described here can be performed in
principle using $n$-photon eigenstates, beamsplitters, phase shifters, and
Kerr nonlinear optical media.  We demonstrate for example how states for our
simplest code may be constructed.  We then discuss how decoding and correction
may be performed.  

\begin{figure}[htbp]
\begin{center}
\mbox{\psfig{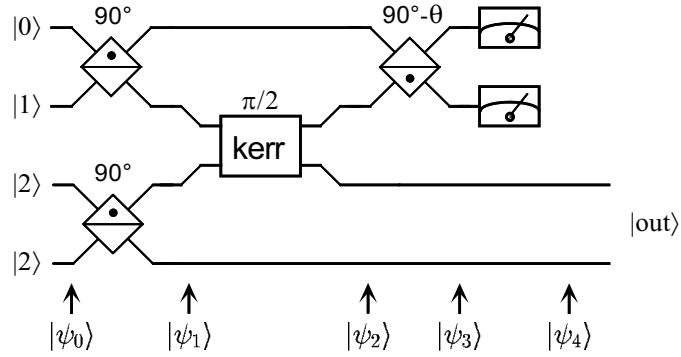}}
\end{center}
\caption{Quantum circuit to encode a qubit using the code of Example~(1).  As
	in \protect\cite{Chuang95}, signals travel from left to right, wires
	represent optical modes, diamonds represent beamsplitters, meters
	ideal photon counters, and the Kerr device is an ideal nonlinear
	optical medium which effects cross-phase modulation of angle $\pi/2$
	between single photons.}
\label{fig:circuit2}
\end{figure}

Shown in Fig.~\ref{fig:circuit2} is a quantum circuit which can be used to
encode a qubit using the $|22\rangle $, $|04\rangle +|40\rangle $ code.  Let
us see how this circuit works by tracing the state at the five indicated
points.  The initial state is $|\psi_0\rangle = |0122\rangle $.  Two 50/50
beamsplitters act on the first and second two modes of this state to give
\begin{equation}
  |\psi_1\rangle  = \left[\rule{0pt}{2.4ex}{ \frac{|10\rangle  + |01\rangle
		}{\sqrt{2}} }\right]  
	    \left[\rule{0pt}{2.4ex}{ \frac{\sqrt{6}(|{04}\rangle+|{40}\rangle)
		- 2 |{22}\rangle}{4} }\right] 
\,.
\end{equation}
Next, a nonlinear optical Kerr medium is used to perform a cross phase
modulation between the two middle modes.  This serves to ``label'' the
$|22\rangle $ state, giving:
\begin{eqnarray}
|\psi_2\rangle  &=& \frac{1}{4\sqrt{2}} \left[\rule{0pt}{2.4ex} {
	    \left(\rule{0pt}{2.4ex}{ |10\rangle +|01\rangle  }\right) \left[{
		\sqrt{6} ( 
		|04\rangle +|40\rangle ) }\right]  
	}\right.
\nonumber \\ && \hspace*{2ex}
	\left.{\rule{0pt}{2.4ex}
	   +  \left(\rule{0pt}{2.4ex}{ |10\rangle -|01\rangle  }\right)
		\left[{ 2 |22\rangle  
		}\right]
		}\right]	
\end{eqnarray}
A final beamsplitter in the first two modes now serves to turn the
phase modulation into a detectable amplitude difference,
\begin{eqnarray}
|\psi_3\rangle  &=& \frac{1}{2\sqrt{2}} \left[\rule{0pt}{2.4ex} {
	    |01\rangle  \left[{ \sqrt{3} \sin\theta
		(|{04}\rangle+|{40}\rangle)  
				+\sqrt{2} \cos\theta |22\rangle  }\right] 
	}\right.
\nonumber \\ && \hspace*{2ex}
	\left.{\rule{0pt}{2.4ex}
	+
	    |10\rangle  \left[{ \sqrt{3} \cos\theta (|{04}\rangle+|{40}\rangle)
				-\sqrt{2} \sin\theta |22\rangle   }\right] 
		}\right]	
\end{eqnarray}
such that if the first two modes are measured to be $|01\rangle $
(otherwise, the state is discarded) then we have the output
\begin{equation}
|\psi_4\rangle  = \cos\theta' |22\rangle  + \sin\theta' \frac{|04\rangle
		+|40\rangle }{\sqrt{2}} 
\,,
\end{equation}
where
\begin{equation}
	\theta' = \tan^{-1}\left[\rule{0pt}{2.4ex}{ \sqrt{\frac{1}{3}}
			\tan\theta }\right] 
\end{equation}
is the new effective angle.  $|\psi_4\rangle $ is the desired encoded state.
After transmission of $|\psi_4\rangle $ through a lossy communication link,
the final state can then be measured immediately (using, for example, photon
number counters) and the transmitted qubit collapsed.  This would be the
standard procedure for point-to-point quantum cryptography.

Alternatively, the qubit may be relayed by performing an error correction
step.  This involves calculation of the error syndrome, correcting any
detected error, then re-encoding the state for further transmission.  In this
system, we have a non-binary state.  If it can be turned into a binary state,
then the entire correction procedure may be performed using standard
techniques of quantum computation, with the usual binary quantum logic
gates\cite{Barenco95}.  This is possible as follows.

\begin{figure}[htbp]
\begin{center}
\mbox{\psfig{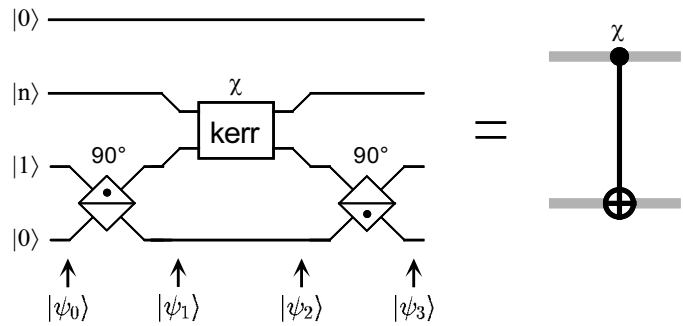}}
\end{center}
\caption{(left) Optical quantum logic gate used as a building block in
	 the decoding procedure; example input states are shown on the
	 left.  $\chi$ is the cross-phase modulation strength of the
	 Kerr medium, which performs the transformation
	 $\protect\exp(i\chi a^\dagger a b^\dagger b)$ on two modes
	 $a$ and $b$ . (right) Shorthand notation for this circuit.}
\label{fig:qfg}
\end{figure}

Consider the circuit shown in Fig.~\ref{fig:qfg}.  The structure is identical
with the well-known quantum-optical Fredkin gate\cite{Yamamoto88,Milb89}, but
let us think of it here in a different way.  The lower pair of wires may be
considered to be a single ``dual-rail'' qubit\cite{Chuang95}, with logical
states $|0_L\rangle = |10\rangle $ and $|1_L\rangle = |01\rangle $.  With an
input state $|\psi_0\rangle = |0n\rangle |0_L\rangle $ and a Kerr medium with
$\chi=\pi/n$, then the output state will be $|\psi_3\rangle = |0n\rangle
|1_L\rangle $, and in this manner the circuit may be thought of as a kind of
{\em controlled-not gate} which distinguishes between a control state of
$|0\rangle $ and $|n\rangle $.

\begin{figure}[htbp]
\begin{center}
\mbox{\psfig{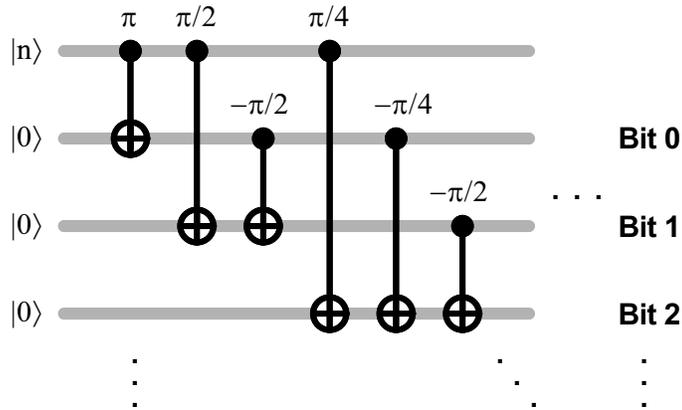}}
\end{center}
\caption{Quantum circuit used to decode a bosonic state into qubits.  Each
	thick wire represents a pair of bosonic modes.  The top wire carries
	the bosonic state, and the remaining wires are prepared as dual-rail
	quantum bits, which carry just one photon in each pair.}
\label{fig:dtob}
\end{figure}

We now use this bosonic controlled-not gate to construct the circuit
of Fig.~\ref{fig:dtob}, which is based on the fact that different
values of $\chi$ allow us to distinguish different values of $n$.
Furthermore, since the decomposition of the number $n$ into sums of
powers of two is unique, it is convenient to take $\chi$ to be binary
fractions of $\pi$.  In this manner, the first dual-rail qubit becomes
the least-significant bit of $n$, and so-on.  If the qubits are
measured, this circuit would then function equivalently to a perfect
photon number detector.  

However, it is much more useful in that calculations may be performed based on
the binary representation of $n$ to determine if any error occurred and to
calculate the error syndrome.  The circuit is then applied in reverse to undo
the entanglement with the bosonic state, and the appropriate correction
procedure is applied to fix the detected error.  This is possible since
generalized measurements may be performed on the qubit states to determine the
error syndrome without destroying the superposition state of the original
qubit encoded in the bosonic state.

\section{Conclusion}

Our treatment of amplitude damping errors is somewhat unusual from the
standpoint of most quantum error correction theories, which deal with a bit
flip and phase flip picture of errors.  The relationship can be understood by
expressing the $A_0$ and $A_1$ operators as coherent superpositions of such
errors; from Eqs.(\ref{eq:adqubit}) and (\ref{eq:adqubitend}),
\begin{eqnarray}
	A_0 &=& \frac{1}{2}\left[\rule{0pt}{2.4ex}
		(1+\sqrt{1-\gamma}) I + (1-\sqrt{1-\gamma}) \sigma_z
		\right]
\\	A_1 &=& \frac{\sqrt{\gamma}}{2} \left[\rule{0pt}{2.4ex}{ \sigma_x +
		\sigma_y }\right]
\,.
\end{eqnarray}
With probabilities up to ${\cal O}(\gamma)$, a binary code with $m$ bits will 
either project $A_0^{\otimes m}|c_l\rangle $ onto a state with no 
errors, or project $A_0^{\otimes m-1} A_1|c_l\rangle $ onto a state with only 
one bit flip error, resulting from a combination of $I^{\otimes m-1}$ and
one of the Pauli operators $\sigma_x$, $\sigma_y$.  Hence, a binary code 
correcting for any one bit error {\em will\/} indeed correct all amplitude 
damping errors up to losing one photon, although not to all orders.  
One reason we have studied bosonic codes
is to exploit the possibilities for achieving higher efficiencies or easier
physical implementation, though the study is theoretically interesting
on its own.

It is important to realize that amplitude damping errors are not independent
bit errors, since the decay factor of each QCS depends on the total number of
excitations in it.  This fact is also pointed out by Plenio {\em
et. al.}\cite{Plenio96a}.  Any code correcting a general one bit error can
do so in the presence of possible amplitude damping, and is capable of 
correcting any one bit error in the projected space in which no loss occurs 
as well as correcting a single photon loss in the absence of any other
errors. Two or more amplitude damping
errors projected into the non-identity space are viewed as separate
errors. The general relation between binary codes and amplitude damping is
still under investigation.

Nevertheless, some interesting comparisons may be made.  Rates from our
bosonic codes contrast with those achievable by the usual binary codes.  For
the code of Example~(1), $nd=4$, $m=2$ and $k=1$, so the rate is found to be
$r =0.22$. This is slightly better than $r=0.20$ for the five bit $(t,k) =
(1,1)$ binary perfect code\cite{Laflamme96}, and much better than the eight
bit $(1,1)$ code of Plenio {\em et. al.}\cite{Plenio96a} which corrects errors
in the presence of $A_k$'s in ${\cal K}(0)$.  Similarly, for the code of
example~(2), $d=2$, $n=6$, $m=3$ and $2^k = 10$ codewords may be found, giving
a rate $r =0.2994$.  In comparison, a naive evaluation of the quantum Hamming
bound\cite{Gottesman96} for binary codes gives a possible rate of $0.41$.
Non-deformation constraints are more restrictive on bosonic error correction
codes than on binary codes, but the bosonic states admit coding schemes which
are impossible with binary codes. There is no conclusive statement on
comparing the general efficiencies of the two different type of codes, but the
examples we have discovered indicate the existence of a rich variety of
bosonic codes which may be useful in the future.

The code fidelities may also be compared.  Our $[[4,2,2,2]]$ code achieves
${\cal F} \approx 1-6\gamma^2$.  In comparison, from an explicit evaluation of
the effect of amplitude damping on all qubits, we have found that the five-bit
$(1,1)$ binary code achieves fidelity $\approx 1-1.75 \gamma^2$, while the
eight-bit $(1,1)$ code achieves only $\approx 1-6\gamma^2$!  This agreement
with the bosonic code is not accidental; it stems from the use of the same
total excitation number.  However, it is worthwhile to point out that despite
the effort to balance the codewords, the five-bit code still has better
performance on average, due to the small number of excitation involved in
the system.

In conclusion, we have given general criteria for an error correction code
which encodes qubits in bosonic states.  This is a generalization of the
binary error correction codes.  Motivated by the dominant decoherence process
(amplitude damping) of system such as photons transmitted through optical 
fibers we classify our errors according to the number of excitations lost, 
instead of the more common classification of the number of bits or modes 
corrupted.  We have shown, in one case, specialization to correction amplitude 
damping does improve the rate, number of encoded qubits to number of required 
qubits. However, bosonic codes under amplitude damping suffer constraints 
involving deformation of the Hilbert space not shared by the binary codes, 
rendering the efficiencies lower in the bosonic case when many qubits are 
encoded.

It is too early to conclude on the relative performance of the binary codes
and the bosonic codes.  Further study will aim at improving the efficiencies,
perhaps by using relative phases to maintain orthogonality of the codewords
instead of by using distinct QCS, so that the QCS can occur in more than one
codeword.  Another possibility is to encode qubits using QCS of different
total number of excitations.  We hope that further study of bosonic codes will
lead to their practical utilization in addition to the current theoretical
interest.

\section{Acknowledgments}

We would like to thank John Preskill for particularly useful discussions.  DWL
was supported in part by the Army Research Office under grant
no. DAAH04-96-1-0299.  ILC acknowledges financial support from the Fannie and
John Hertz Foundation.

 


\appendix

\section{Criteria for non-deformation of Hilbert Space}

Consider ${\cal K}(2)$ = $\{0\cdots02, 0\cdots20,\ldots, 20\cdots0\}$ 
$\bigcup$ $\{0\cdots011, 0\cdots110, \cdots, 110\cdots0\}$ for $t=2$
errors. For instance,
\begin{eqnarray}
	& & A_{0\cdots 02} |c_l\rangle 
\\	&=& A_{0\cdots 02} \frac{1}{\sqrt{N_l}}
				\sum_{i=1}^{m} |n_{i1}\ldots n_{im}\rangle 
\\
	&=& \sum_{i=1}^{m} 
		\sqrt{\frac{{\cal C}(n_{im},2)\gamma^2 (1-\gamma)^{N-2}}
			   {N_l}}
		|n_{i1}\ldots n_{im}-2\rangle 
\,,
\end{eqnarray}
where ${\cal C}(n,m)$ is the usual binomial coefficient.  The norm square
of this state is
\begin{equation}
	\langle c_l| A_{0\cdots02}^\dagger A_{0\cdots02} |c_l\rangle  
	= \frac{\gamma^2 (1-\gamma)^{N-2}}{2N_l}
		\sum_{i=1}^{m} n_{im}(n_{im}-1)
\,.
\end{equation}
The term linear in $\gamma$ is independent of $l$ (codeword independent) by
the criteria for $t=1$; hence, it follows that:
\begin{equation}
	\frac{1}{N_l} \sum_{i=1}^{m} n_{im}^2
\,
\end{equation}
has to be independent of $l$ if the non-deformation criteria is to be
satisfied.  Other $A_{\tilde{k}}$ with ${\tilde{k}}=0\cdots02,
\ldots, 20\cdots0$ impose the above requirement on other columns.

Similarly, ${\tilde{k}}=0\cdots11$ changes the codeword to:
\begin{equation}
	\sum_{i=1}^{m} 
	  \sqrt{\frac{n_{im-1}n_{im}\gamma^2 (1-\gamma)^{N-2}}{N_l}}
  	      |n_{i1}\ldots n_{im-1}-1 n_{im}-1\rangle 
\,.
\end{equation}
which has norm square 
\begin{equation}
	\frac{\gamma^2 (1-\gamma)^{N-2}}{N_l} \sum_{i=1}^{m} n_{im-1}n_{im}
\,.
\end{equation}
Eq.(\ref{eq:crittwo}) requires the following to be independent of $l$:
\begin{equation}
	\frac{1}{N_l} \sum_{i=1}^{m} n_{im-1}n_{im}
\,.
\end{equation}
A similar result is obtained for other ${\tilde{k}}$ with 1's at any two modes
$j_1$ and $j_2$. When we allow $j_1=j_2$, we include the previous result for
two photon loss at one mode.  This proves {\em Lemma~2}.

For arbitrary $t$, we get equations involving various binomial coefficients.
Using requirements involving products of fewer than $t$ $n_{ij}$'s, we can 
replace the products of the binomial coefficients to products involving
exactly $t$ $n_{ij}$.  By mathematical induction, the result for arbitrary 
$t$ is then obtained. This completes the proof of {\em Theorem~1}.

\end{multicols}


\clearpage

\end{document}